\documentclass[epj]{svjour}
\usepackage{graphics}
\usepackage{amssymb,amsmath,mathrsfs,tikz}
\usepackage{subfig}
\usepackage{graphicx,caption,float}
\usepackage{color}
\usepackage{gensymb}
\usepackage{hyperref}
\usepackage{listings}
\usepackage[utf8]{inputenc}
\begin{document}
\title{Topology and ground state degeneracy of  tetrahedral smectic vesicles}
\author{Francesco Serafin\inst{1,2}  \and Mark J. Bowick\inst{2} \and Sidney R. Nagel\inst{3}
\thanks{\emph{Present address:} fserafin@syr.edu, srnagel@uchicago.edu, bowick@kitp.ucsb.edu}%
}                     
%
%
\institute{Physics Department and Syracuse Soft and Living Matter Program,
Syracuse University, Syracuse, NY 13244, USA
\and Kavli Institute for Theoretical Physics, University of California, Santa Barbara, CA 93106, USA.
\and The James Franck and Enrico Fermi Institutes and The Department of Physics,
The University of Chicago, Chicago, IL 60637, USA}
\date{Dated: July 15, 2018. Revised version: October 4, 2018. }
%
\abstract{
Chemical design of block copolymers makes it possible to create polymer vesicles with tunable microscopic structure. Here we focus on a model of a vesicle made of smectic liquid-crystalline block copolymers at zero temperature. The vesicle assumes a faceted tetrahedral shape and the smectic layers arrange in a stack of parallel straight lines with topological defects localized at the vertices. We counted the number of allowed states at $T=0$. For any fixed shape, we found a two-dimensional countable degeneracy in the smectic pattern depending on the tilt angle between the smectic layers and the edge of the tetrahedral shell. For most values of the tilt angle, the smectic layers contain spiral topological defects. The system can spontaneously break chiral symmetry when the layers organize into spiral patterns, composed of a bound pair of $+1/2$ disclinations. Finally, we suggest possible applications of tetrahedral smectic vesicles in the context of functionalizing defects and the possible consequences of the spiral structures for the rigidity of the vesicle.
\PACS{{}{}
     } 
} 
\maketitle
\section{Introduction}
\label{intro}
From the largest cosmic structures down to the atomic scale, we are surrounded by a rich variety of shapes. Although shape is often the first feature that we identify in a physical system, understanding its origin is in most cases difficult and challenging. Morphogenesis \cite{morphogenesis} typically requires the collective organization of a macroscopically large number of degrees of freedom, either in or out of thermodynamic equilibrium, moreover the concept of shape itself is often scale-dependent. Here we are interested in characterizing shapes of systems that are large compared to their elementary constituents. For this reason, we will work within a coarse-grained, continuum field-theoretic framework \cite{book:ZinnJustin}, since we expect that the observables of interest will depend only on universal properties of the system and not on its microscopic details.  
\par In this work we consider a closed vesicle of spherical topology formed by spontaneous aggregation of smectic liquid-crystal molecules (a common example is the rod-coil block copolymer structure) \cite{B902003A,C7PY01131H,B907485F,C6PY00545D}. The smectic liquid crystal organizes in a layered structure \cite{book:DeGennes}, which we idealize as a set of parallel lines on the vesicle. Because they are very flexible and thin, vesicles are interesting systems in which to study morphogenesis \cite{doi:10.1080/00018739700101488,WINTERHALTER199335}. Their thickness is of the order of nanometers, while their size can be of the order of 100 micrometers making them a good physical realization of a two-dimensional fluid membrane \cite{book:ZinnJustin}. We modeled the vesicle shape as a 2-dimensional manifold $\mathbf{X}$ embedded in 3-dimensional space, 
while the smectic order parameter is described by a 1-form field $\omega$ defined on it. 
\par Consider an ensemble of such vesicles. Working in the limit of extremely soft (easily bendable) surfaces, we expect to encounter a large number of low-energy shapes, leading to a large entropic contribution. To each configuration we associate a free energy $F[\mathbf{X},\omega]$ that measures the energetic cost of bending the surface and of deforming the liquid-crystal layers. The presence of the liquid crystal and its elastic response to deformations has a strong influence on the equilibrium shapes.
\par We studied the ground state of $F$, in other words the density of states at zero temperature. This is in general a very hard infinite-dimensional minimization problem, because both the manifold $\mathbf{X}$ and the field  $\omega$ are allowed to vary. This situation is reminiscent of General Relativity with matter fields, or of classical solutions to some models of two-dimensional quantum gravity \cite{1987gauge}, where one solves simultaneously for the manifold geometry as well as for the matter fields.  We can make a step towards the solution when we notice that the vesicle's spherical topology forces the smectic field to contain topological defects at isolated points, even in the ground state. 
\par A sphere has an Euler characteristic equal to 2, so by the Gauss-Bonnet theorem \cite{book:docarmo} the total topological charge of the smectic field must add to 2.
When the system's shape is allowed to change, and in the absence of topological obstructions \cite{BALPhysRevLett.52.1818,baues2006obstruction,husemoller1994fibre}, the coupling between defects and curvature determines both the equilibrium shape and the smectic configuration. For an extensive review on the interplay of curvature, order and topological defects see \cite{2009AdPhy..58..449B}. Recent work \cite{Xing5202} showed that the ground states of $F$ span a two-dimensional manifold of faceted tetrahedral shells. At a first glance, it seems surprising that, just by coupling to the liquid-crystal order, a very soft and fluid system can develop sharp edges and vertices. One equilibrium smectic pattern was found to be a set of lines parallel to an edge of the tetrahedron, with four disclinations of charge 1/2 located at the polytope's vertices.
\par In this work we considered a tilt angle between the smectic layers and the edge of the tetrahedral shell and characterized the degeneracy of such ground-state manifolds. Many of these states contain spiral defects. A complete classification of smectic defects can be found in \cite{Xing2009}. It is interesting to note that the smectic vesicle realizes all of them in the ground state. 
\par The lowest-energy manifold displays a number of interesting features:
First, the presence of spiral states signals the emergence of chirality in the system composed of achiral building blocks.
Second, a smectic state breaks translational invariance to a discrete subgroup. As a consequence, a state exists only if the layer separation $h$ and the vesicle size $L$ are commensurate. For special values of the ratio $L/h$, a fixed shape supports more than one tilt angle and therefore many distinct topological classes of defects. 
\par In some states the smectic layers are arranged as the latitudes on a globe. In this case, since there is no flow of molecules across layers, a localized perturbation of the fluid within a layer must remain confined on a fixed latitude. In contrast, we expect a spiral state to be much less rigid because a localized deformation would be able to propagate along the spiral and travel long distances from the initial perturbation.
\par We imagine that smectic vesicles could be used experimentally to self-assemble micro-scale sacks. If the defect sites could be functionalized by anchoring ligands to the defect cores \cite{2002NanoL...2.1125N,DeVries358}, these systems would display highly anisotropic interactions, and the presence of chiral spiral patterns might perhaps open up interesting possibilities in the context of supramolecular chemistry \cite{akashi2000giant}. 
\section{The model}
\begin{figure}
\includegraphics[width=9.5cm]{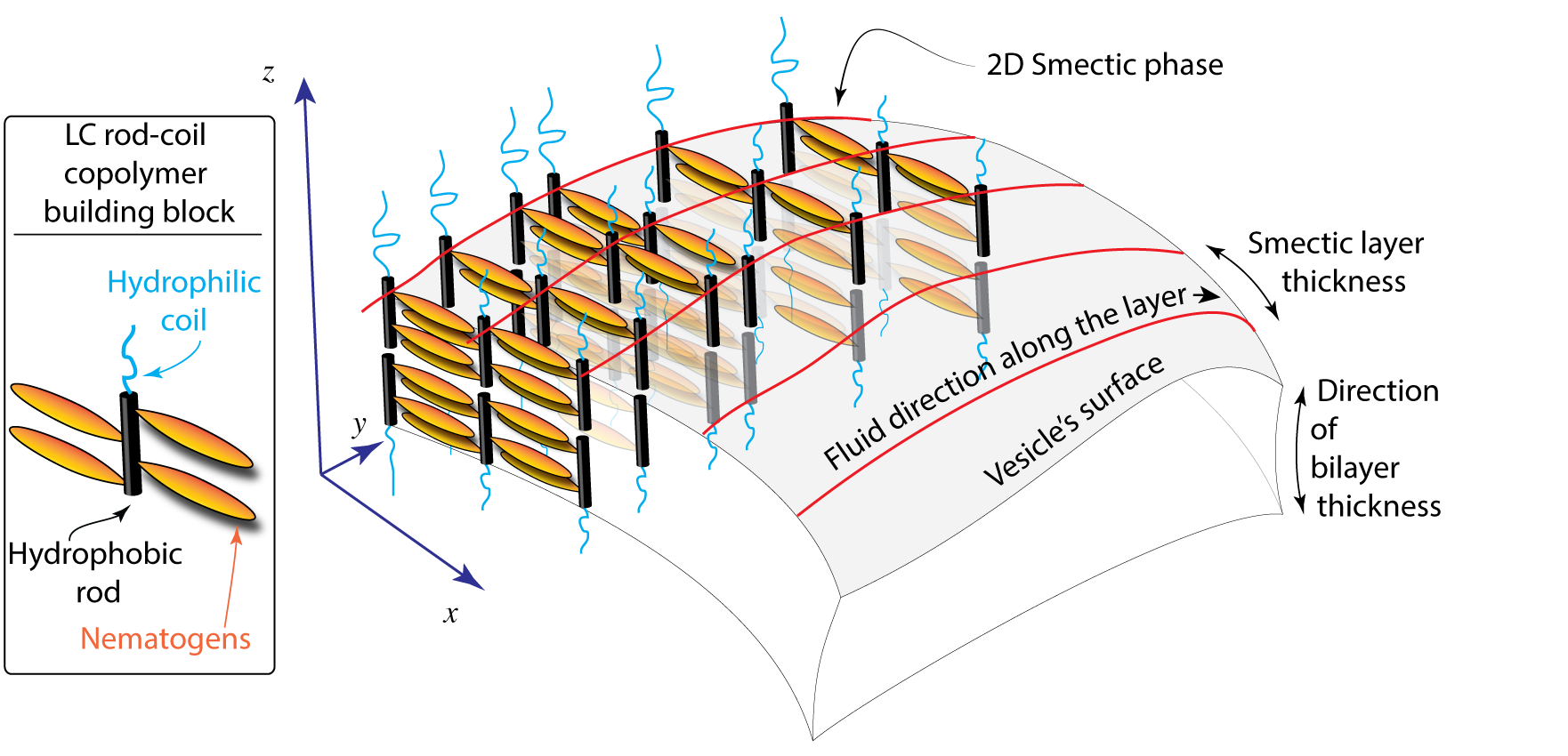}
\caption{\label{fig:SM_membr}(\emph{Left})   { }{Schematic of the molecular building blocks that form the vesicle. (\emph{Right}) Section of an LC membrane with 2-dimensional smectic-A ordered phase on its tangent plane. Notice the interdigitation of the   { }{mesogens} along the local $x$ direction, as well as their alignment along the $y$ direction.}}
\end{figure}
As mentioned in the introduction, the molecular constituents of the vesicle behave as a liquid crystal. 
  { }{ Recently it was shown that certain rod-coil copolymers with side   { }{mesogens}  can spontaneously self-assemble into bilayered membranes in an aqueous environment \cite{B902003A,C7PY01131H}. {A solution made of the side   { }{mesogens} alone is known to possess a smectic bulk phase.} fig. \ref{fig:SM_membr} (\emph{Left}) shows a schematic of these elementary building blocks. Reference \cite{C7PY01131H} reported that self-assembled liquid crystalline vesicles may have smectic order. Fig. \ref{fig:SM_membr} (\emph{Right}) shows how the rod-coil copolymers are thought to be arranged to make the membrane. We choose the $z$ direction to be aligned with the rods, i.e. with the normal to the membrane. The $x$ and $y$ directions span the surface of the vesicle. We are interested in the ordered structure in the local tangent plane $xy$.
Consider the plane $xz$ at $y=0$.  The rod-coil copolymers self-organize in a bilayer structure where the   { }{mesogens} are locked in an inter-digitated pattern. The rod is normal to the membrane. As we move along the y axis, we  find a sequence of stacks containing the same pattern as in the $y=0$ plane.
The key point is that the   { }{mesogens} (orange in the picture) interact as a smectic liquid crystal along the $y$ direction too. So the   { }{mesogens}' center of mass in the second layer to the right of the plane $y=0$ must be aligned with the   { }{mesogens}' center of mass in the $y=0$ layer.  The same happens for the third layer along the $y$ axis, and so on.  The   { }{mesogens} organize in a layered pattern on the $xy$ plane (the vesicle's surface) where the layer normal coincides with the   { }{mesogens}.  This results in a sequence of stripes on the surface of the vesicle (i.e. the $xy$ plane) that breaks translational symmetry along the $x$ direction \cite{C7PY01131H}. The layer spacing has the same value as in the smectic bulk phase of the nematogens alone. In Ref. \cite{Xing5202} the smectic structure on the surface is described as a nematic director field where the in-plane bending constant is large compared to the in-plane splay constant. The positions of the black rods along the $y$ axis is arbitrary, so the membrane is fluid along this direction. }
Following \cite{Xing5202}, we consider a minimal model of a liquid crystal coupled to a curved manifold $M$ of spherical topology
\footnote{By energetic considerations, a large vesicle inhibits the formation of holes. The energy cost for cutting a hole grows proportionally to the hole's boundary, with the line tension as a proportionality factor.  In the thermodynamic limit, the energy of the boundary diverges. On the other hand, the bending energy remains constant as we increase the system size, because it is conformally invariant \cite{2012arXiv1202.6036M,MR0324603}. So we expect that shape fluctuations will close the holes and self-heal the surface, reducing it to a spherical topology. }.  
Assuming that the vesicle is formed by two symmetric monolayers, the free energy is
\begin{equation}
\label{eq:F}
F[\mathbf{X},\mathbf{n}]=\int_M \sqrt[]{g} \,d^2x \,[K_1(\nabla\cdot \mathbf{n})^2+K_3(\nabla\times \mathbf{n})^2+\kappa H^2] 
\end{equation}
where $\mathbf{n}$ is the director field associated to 
the liquid-crystal molecules, $\mathbf{X}$ is the embedding of $M$ in $\mathbb{R}^3$, $g=\sqrt[]{\det g_{\alpha\beta}}$ is the determinant of the induced metric on $M$,  $\nabla$ is the covariant derivative on $M$, $H$ is the mean curvature \cite{book:docarmo}, $K_1$ and $K_3$ are respectively the splay  and bending moduli of the liquid crystal, and $\kappa$ is the bending rigidity of the system. Notice that the surface tension $\mu$ is set to zero. We chose a thermodynamic ensemble where the vesicle is coupled to a reservoir of liquid-crystal molecules. The area can freely fluctuate as molecules flow on or off the surface, so $\mu\equiv\delta F/\delta A =0$. We study the ensemble at zero temperature, in other words we look for minimizers of $F$.
\par  The functional \eqref{eq:F} exhibits an interesting set of degenerate ground states. We will study \eqref{eq:F} in the limit $K_3/K_1\to\infty$, when the field $\mathbf{n}$ becomes constrained by $\nabla\times \mathbf{n}=0$, and describes a smectic liquid crystal. In this limit the field configurations contain pure splay modes. The integral lines of the field $\mathbf{t}$ orthogonal to $\mathbf{n}$ will be interpreted  as the smectic layers. Equivalently, we can say that they are described by the level sets of a 1-form $\omega$ (see next section). In addition, we want to tune the bending rigidity $\kappa$ to zero, where the membrane is extremely soft. Then the energy density reduces to $K_1(\nabla\cdot\mathbf{n})^2$, and the equations of motion are $\nabla_a(\nabla_bn^b)=0$, subject to the constraint $\int_M K_G = 2\pi\chi=4\pi$ , where $K_G$ is the Gaussian curvature and $\chi$ is the Euler characteristic. 
\par  If we ignored the constraint, we would conclude that \eqref{eq:F} is minimized by a flat manifold ($\nabla=\partial$) covered by a uniform field of straight lines ($\partial^2\mathbf{n}=0$). However, this cannot be true when we require that the manifold retain its spherical topology. In order to satisfy the condition $\chi=2$, we conclude that $M$ must be a developable surface ($K_G=0$ almost everywhere) and that the director field $\mathbf{n}$ follows parallel straight lines on $M$, except for isolated singular points. 
{It follows that the smectic layers,  represented by the integral lines of $\mathbf{t}$, where $\mathbf{t}\cdot\mathbf{n}=0$, are also a set of parallel straight lines. } The gaussian curvature is localized in a discrete set of points $S$, identified with the vertices of a faceted  polyhedron, and it is screened by the topological defects of the field $\mathbf{n}$. 
\par  The minimal number of points that span a non degenerate polyhedron is 4, so the director field will contain four disclinations of charge $s=1/2$, and the ground state shape will be a tetrahedron \cite{lubenskyProst} (see also \cite{PhysRevLett.99.157801,FrustrOrder} for the nematic analogue). 
The region of parameter space that we want to study ($K_3/K_1\to\infty,\,\kappa=0$) allows many distinct ground state configurations. The analysis developed in \cite{Xing5202} showed that the ground state tetrahedra span a two-dimensional moduli space. For a fixed shape, there are many configurations of the field $\mathbf{n}$ (and of the corresponding smectic field) that lie at the same energy. In the following sections we will address the problem of characterizing the degeneracy of these smectic field configurations.
\section{\label{sec:two}Geometry of smectic layers}
Smectic liquid crystals are composed of apolar elongated molecules that, below a critical temperature, exhibit an ordered phase with broken orientational and translational symmetry \cite{book:DeGennes,Kleman}. A smectic liquid crystal can break orientational symmetry thanks to local interactions that favor alignment of the molecules. In addition to this nematic order, a smectic state also forms layers, so that it also breaks continuous translational invariance to a discrete subgroup of translations along a direction that coincides (locally) with the average molecule orientation axis. In a smectic-A liquid crystal, within each layer the average molecular orientation coincides with the local layer's normal. 
\par We will consider 2-dimensional closed membranes made of self-assembled molecules that behave as an incompressible smectic liquid crystals on the tangent space of the surface. We will adopt a coarse-grained point of view and parametrize the liquid-crystal vesicle as a 2-dimensional manifold $M$ embedded in $\mathbb{R}^3$ with induced metric $g_{\alpha\beta}$. We will call $S$ the set of points that host topological defects. The smectic layers are then described by a measured foliation \cite{Poenaru1981,book:Lawson} of $M\smallsetminus S$ (where the singular points have been removed), that is, a collection of disjoint connected curves on $M\smallsetminus S$ that are equispaced and that cover $M\smallsetminus S$.
The direction of the molecules defines locally a vector field $\mathbf{n}(x)$. We associate to it a 1-form $\omega$ such that 
\begin{equation}
\omega[\mathbf{n}]\equiv1 \quad.
\end{equation} 
A smectic layer is a curve $\mathbf{r}(s)$ on $M$ with tangent $\mathbf{t}(s)$, where $\mathbf{t}$ is the solution to the equation $\omega[\mathbf{t}]=0$.
The smectic layers are separated by a distance $h$, which is uniform on $M$. This implies that the contraction of $\omega$ with an arbitrary vector is an integer multiple $q\in\mathbb{N}$ of $h$: $\omega[\mathbf{v}]=q h$. Notice that, because $\omega$ is defined locally, a smectic layer that bends over itself counts as two layers. 
  { }{There exist generalizations to this simple level set construction that describe equispaced smectic patterns in a $d$-dimensional curved space as level sets of null hypersurfaces in $d+1$ dimensional Minkowski, de Sitter
and Anti-de Sitter spacetimes \cite{MinkowskiSmPhysRevLett.104.257802,ConformalSmectic}. In that formulation, distinct smectic configurations are related by coordinate transformations.}\section{\label{sec:three}Structure and energetics of defects in smectics}
The orientational order parameter associated to the smectic liquid crystal is a nematic-like director field. Topological defects (disclinations) in the orientation of the smectic liquid crystal appear at isolated points, and their charge is an integer multiple of 1/2 \cite{MICHELRevModPhys.52.617,Mermin_RevModPhys.51.591}. We usually detect their charge by encircling the defect by a closed path, and counting how many times the director rotates back to itself as we complete a revolution around the defect. Every time the director rotates by $\pi$, we have closed a loop in the order parameter space \cite{kleman1977classification}. The lowest disclination charge is therefore $q=\pm\pi/(2\pi)=\pm1/2$. Because the disclination energy is proportional to the square of its charge, both in a flat and in a curved surface, the elementary disclinations with $q=1/2$ are favored over a $q=+1$ defect \cite{lubenskyProst}.
\par
\begin{figure}
\includegraphics[height=5.5cm]{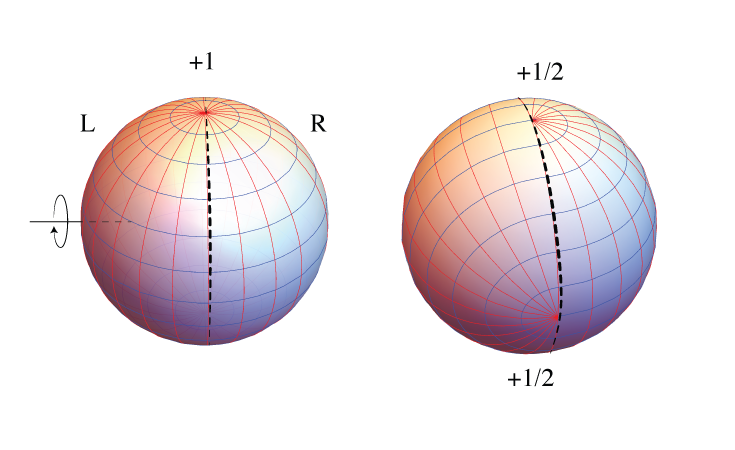}
\caption{\label{def_split}Splitting a +1 defect into two 1/2. Cut the sphere  along the dotted line (left) and rotate the left hemisphere until the smectic layers (blue lines) are reconnected.}
\end{figure}

 The classification of the field configurations requires knowing the relative energies between the allowed classes of topological defects. Unlike a nematic, one smectic defect of charge $s=1$ and two defects of charge $s=1/2$ have the same energy. The proof is contained in~\cite{PhysRevLett.101.037802} and~\cite{Xing2009}: parametrize the sphere with angular coordinates $(\varphi,\theta)$ and consider the smectic state $\omega=d\theta$ (Fig.~\ref{def_split}). The smectic layers are the lines $\theta=\pi/N$, where $N$ is the total number of layers. Each layer is a curve of constant latitude. The poles contain two defects of charge +1. Now cut the sphere along the meridian $\varphi=0$ and divide it in right and left hemispheres.
We can rotate the left hemisphere around the $y$ axis by an integer multiple of $\pi/N$ and split each +1 defect into a pair of 1/2 defects. Notice that the integral lines of $\mathbf{n}$ are again segments of geodesics, and they have zero bending everywhere, so they still describe a smectic state. Since the two hemispheres slid along a geodesic, the energy of the configuration does not change. Although these two states are degenerate in energy, only the one that contains four disclinations is compatible with a faceted tetrahedron. The four defects experience a repulsive effective interaction, so in the ground state they maximize their pairwise geodesic distance by occupying the vertices of a regular tetrahedron \cite{lubenskyProst}. 
\par There is another smectic defect whose charge is $q=+1$ but is topologically distinct from the latitudinal state: it is a single spiral contained between the North and the South pole of the sphere (see the classification of smectic defects in \cite{Xing2009}). The spiral defect can be thought of as a closely bound pair of two $+1/2$ disclinations. The energy of this topological defect on a smooth surface of spherical topology ($\kappa\neq0$) is higher than the latitudinal state, because the geodesic curvature of the director's integral lines cannot vanish everywhere. On a faceted shell however ($\kappa=0$), the integral lines become rectilinear segments, and the single spiral state becomes degenerate in energy with respect to the latitudinal state. This shows once again that setting $\kappa=0$ is different from taking the limit $\kappa\to0$.
\section{\label{sec:four}Ground state degeneracy}
\begin{figure}
\includegraphics[height=3cm]{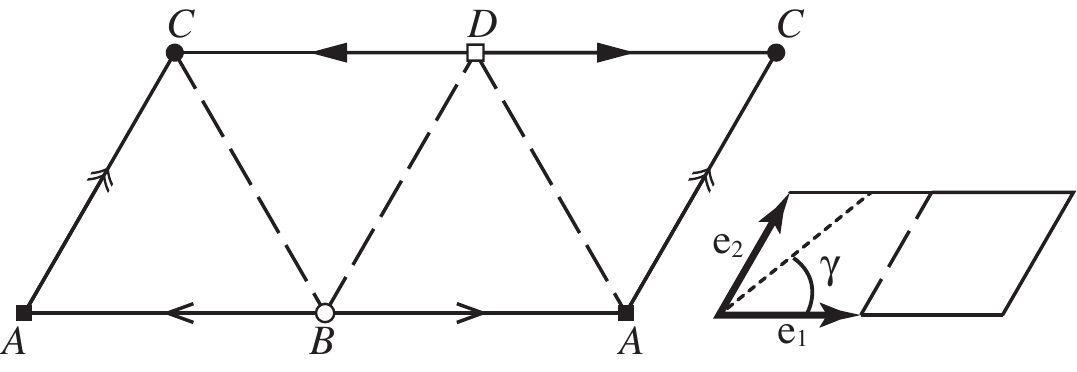}
\caption{\label{fund_dom}\emph{(Left)} The domain $\mathcal{P}$ foldable into a regular tetrahedron. The edges identifications are indicated by arrows and the vertices by letters. The dashed lines indicate the folds. \emph{(Right)} The lattice vectors and the direction $\gamma$ between a layer and the vector $\mathbf{e}_1$}
\end{figure}

Recall that the gaussian curvature is localized at the vertices of a developable tetrahedron where it is screened by the charge of the topological defects. Since the covariant derivative $\nabla\mathbf{n}$ must vanish everywhere except at the vertices, the sum of the angles between the edges around each vertex must be $\pi$, otherwise in the developed tetrahedron the smectic lines would bend to fill uniformly the surface. This constraint means that the only those tetrahedra that are developable into a flat parallelogram $\mathcal{P}$ are allowed, and that all the 4 faces must be identical \cite{Xing5202}. For later convenience, let us call $\mathbf{e}_1$ and $\mathbf{e}_2$ two vectors that span half of the developed parallelogram (Fig. \ref{fund_dom}).  {We are interested in the case $\kappa\equiv0$, where the degeneracy of shapes is maximal. An arbitrarily small deviation of $\kappa\sim0^+$, however, will select the shape that minimizes the total length of its edges (the bending energy is proportional to the length of a rounded edge) which is a regular tetrahedron.} Therefore, for the purpose of classifying states, we will use the regular tetrahedron as the manifold's shape. The same logic applies to non-regular tetrahedra. Notice that the domain $\mathcal{P}$ has the topology of the sphere. The oriented segment $\vec{AC}$ on the left in Fig.  \ref{fund_dom} is identified with $\vec{AC}$ on the right, while the segment $\vec{CD}(\vec{AB})$ is identified with its \emph{reflected} image $\vec{DC}(\vec{BA})$ across the vertex $D(B)$.
\par
\begin{figure}
\includegraphics[width=9cm]{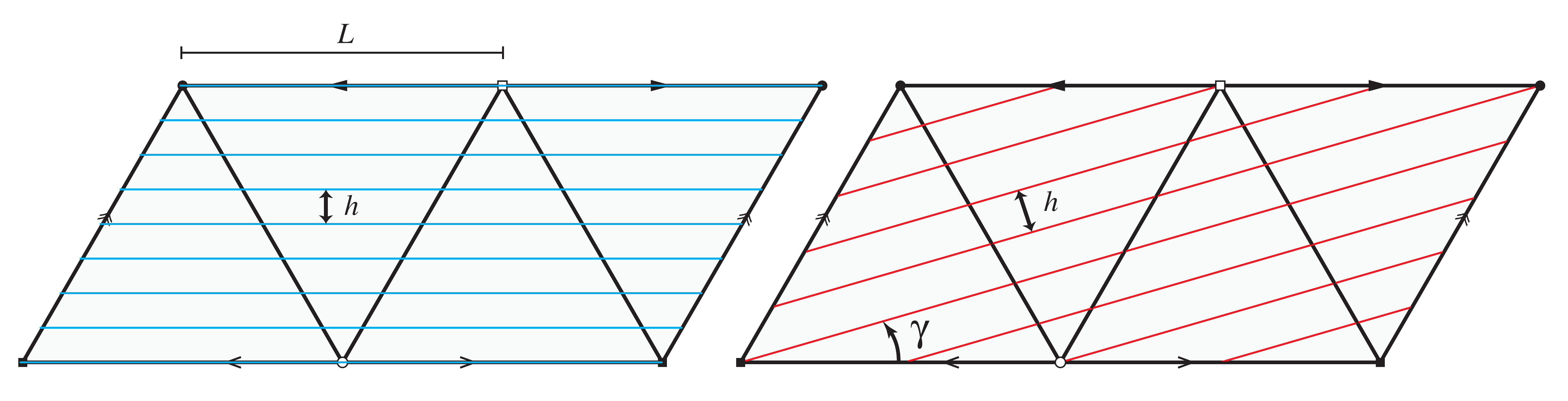}
\caption{\label{tilt} (\emph{Left}) Developed tetrahedron of size $L$ covered by a commensurate uniform smectic pattern with layer spacing $h$. (\emph{Right}) A vesicle where the layers have been tilted by an angle $\gamma$.}
\end{figure}

 Once we fix $\mathbf{e}_1,\mathbf{e}_2$, and the angle between them, we are left with the degeneracy in the configurations of the field $\mathbf{n}$. On a generic surface $M$, the pure-splay constraint $\epsilon^{ab}\nabla_a n_b=0$ (where $\epsilon^{ab}$ is the Levi-Civita symbol) implies 
\begin{equation}
\label{Eq:curl}
\partial_1n_2=\partial_2n_1 \quad.
\end{equation} 
On the other hand, $n_\alpha n^\alpha=1$ and $\partial_\beta(n_1^2+n_2^2)=0$, we find
\begin{align}
\label{eq:norma}
&n_1\partial_1n_1+n_2\partial_1n_2=0\\
\label{eq:normb}
&n_1\partial_2n_1+n_2\partial_2n_2=0 \quad,
\end{align}
where we used the relation $\epsilon^{ab}\Gamma^c_{ab}=0$ thanks to the symmetry of the Christoffel symbols $\Gamma_{ab}^c=\Gamma_{ba}^c$.
Combining \eqref{eq:norma} and \eqref{eq:normb} with \eqref{Eq:curl}, we find 
\begin{equation}
\label{eq:eom}
(n^a \nabla_a)\mathbf{n}=0 \quad,
\end{equation}
that is, the integral lines of $\mathbf{n}$ are geodesics on $M$. Since the tetrahedron is developable into a flat domain $\mathcal{P}$, the integral lines of $\mathbf{n}$ are simple straight lines on $\mathcal{P}$. The smectic layers will also be parallel straight lines, and additionally they must be equi-spaced. 
Most importantly, the equations of motion $\partial^2\mathbf{n}=0$ and the energy are left invariant by a global rotation of the layers, so we expect to find a degeneracy of smectic configurations, labeled by the angle $\gamma$ between a layer and the vector $\mathbf{e}_1$ (Fig. \ref{tilt}). 
\section{Topological classification of the ground states}
We are now going to classify the allowed smectic states by drawing suitable sets of parallel lines on $\mathcal{P}$.  We must follow two rules when we draw the states: (i) each vertex must be occupied by a smectic disclination, and (ii) the layers are equispaced. Condition (i) is a consequence of the structure of a defect with charge 1/2, and eventually determines the allowed orientations $\gamma$ of the layers.
\par
\begin{figure}
\includegraphics[height=4.5cm]{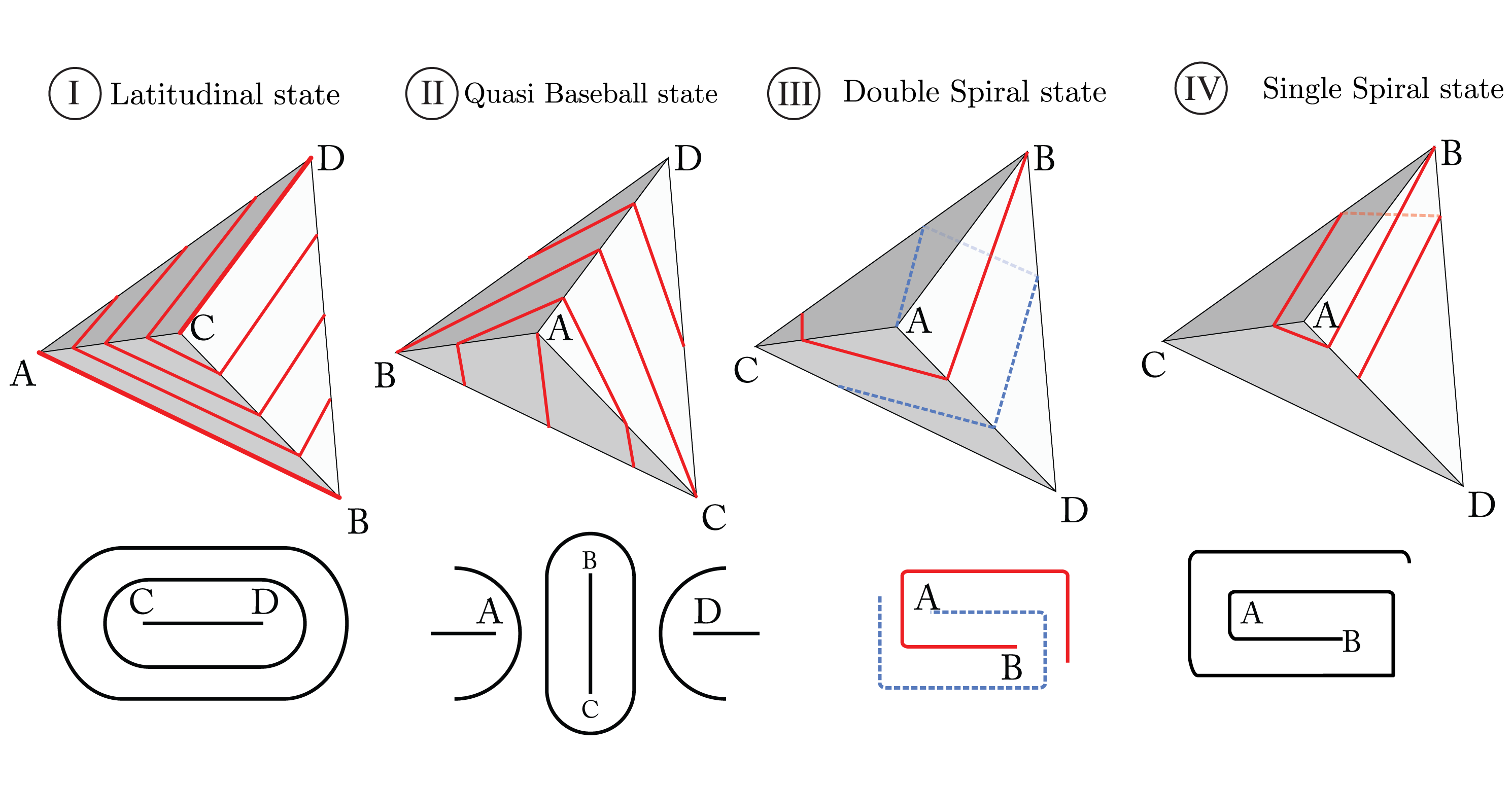}
\caption{\label{classes}
Topological classes of smectic states on the vesicle, and their projection onto a plane. {The states $I$ and $II$ are topologically equivalent (same topological class), but they are distinct with respect to the value of the tilt angle.}  }
\end{figure}

 The easiest state that can be drawn is a set of lines parallel to $\mathbf{e}_1$ $(\gamma=0)$ and spaced by $\sqrt[]{3}/(2N)$, where $N$ is the number of lines (Fig.~\ref{tilt}~\emph{Left} and~\ref{classes}-I). 
This is called a Latitudinal state \cite{Xing5202}, and it has the same topological structure of the state $\omega=d\theta$ on the sphere. The number of independent layers is $\nu=N$. If the edge of the vesicle has length $L$ and the distance between two neighboring smectic layers is $h$, then we have
\begin{equation}
\label{eq:NUM1}
h=\frac{\sqrt[]{3}}{2}\frac{L}{N}
\end{equation}
where $N$ is the number of layers on the vesicle.
\par The next state that we will discuss occurs at $\gamma=\pi/6$. By rule (i) there must be at least two independent layers, one connecting $A$ with $D$ and the other connecting $B$ with $C$. According to rule (ii), an arbitrary number of layers can be inserted between the first two, as long as they are equispaced. The resulting state is the analogue of the quasi-baseball state on the sphere (Fig. \ref{classes}-II). 
 {Although this layer pattern has the same topology as the $\gamma=0$ state, we regard them as different states with respect to their tilt angle with the egdes. 
}  
The layers are perpendicular to an edge. We can either have a layer terminating in the vertices ($L=nh$ with $n$ even) or we can shift the former by $h/2$ and have $L=nh$ with $n$ odd.
The latitudinal and the quasi-baseball states are much less frequent than the next set of states that we are going to discuss.
\par Consider now an angle $\gamma\neq 0, \pi/6$. We will show later that the wedge $\gamma\in[0,\pi/3]$ contains all the states. Following rules (i) and (ii), we cover the domain $\mathcal{P}$ with a measured foliation of parallel lines such that each vertex is occupied by a smectic layer termination.
Because of the topology of $\mathcal{P}$, the smectic layers will connect across the edges to form spirals. We can either form a double spiral (Fig. \ref{classes}-III) or a single spiral (Fig. \ref{classes}-IV).  This means that these states are organized in two classes according to the spiral's 
chirality: Left-handed if $\gamma\in(0,\pi/6)$, and Right-handed if $\gamma\in(\pi/6,\pi/3)$. Notice that right- and left- states are in 1:1 correspondence, so we just need to classify, say, the left-handed states i.e. $\gamma\in(0,\pi/6)$.
For convenience, the handedness of the spirals is best seen by looking at an axis that goes through the mid point of two opposite edges (e.g., $AB$ and $CD$ in Fig. \ref{classes}-III). 
\begin{figure}[b]
\includegraphics[height=2.5cm]{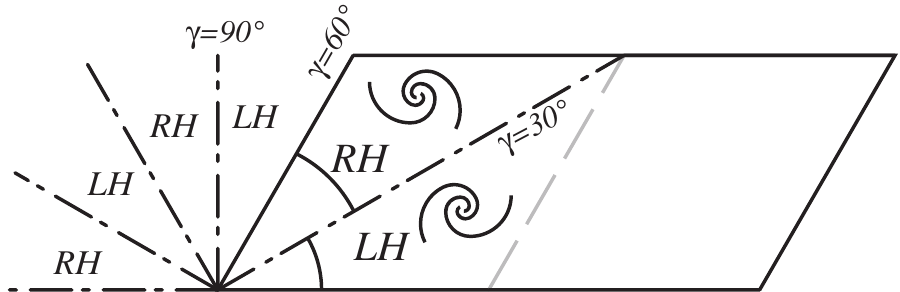}
\caption{\label{angles}The redundancy in the tilt angle $\gamma$: the states are labeled by $\gamma$ mod $\pi/3$. And they split between Left- and right-handed according to $0<\gamma<\pi/6$ or $\pi/6<\gamma<(2\pi)/6$. }
\end{figure}
\par
\begin{figure}[t]
\includegraphics[height=3cm]{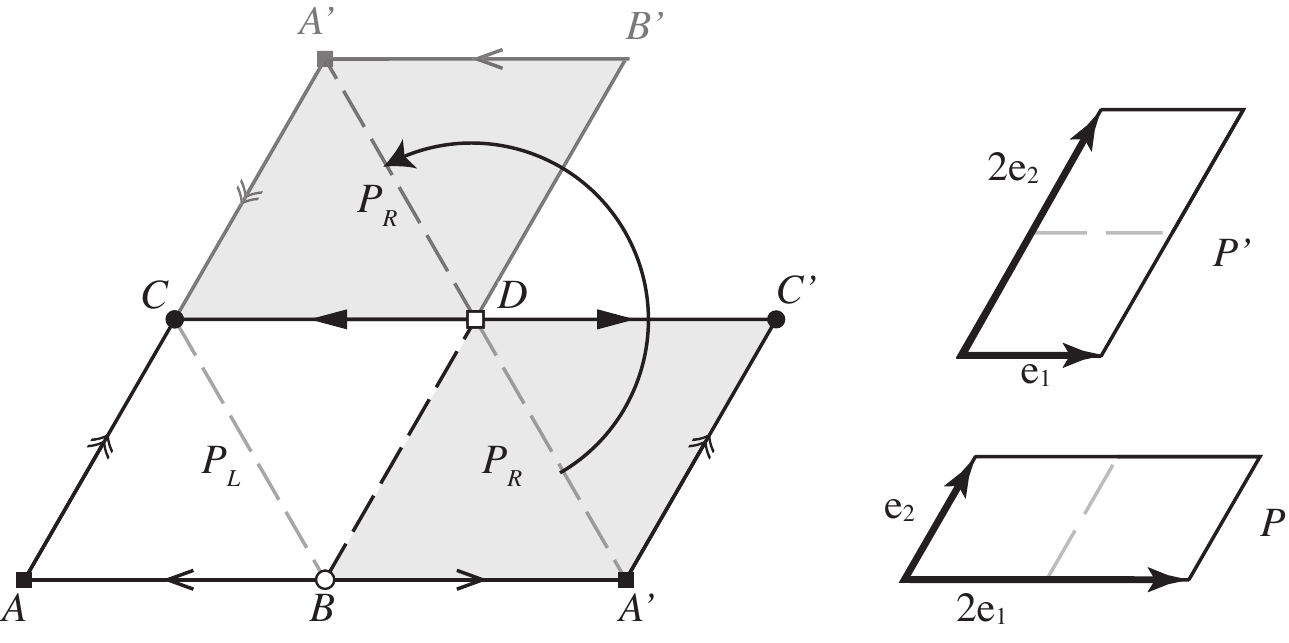}
\caption{\label{domains}Invariance of the fundamental domain under rotations and translations.}
\end{figure}

 To proceed, consider a state with orientation $\gamma$ on $\mathcal{P}$ and divide the domain $\mathcal{P}$ symmetrically into two sub-domains $\mathcal{P}_R,\mathcal{P}_L$ spanned by $\mathbf{e}_1$ and $\mathbf{e}_2$ (Fig.~\ref{domains}). 
  { }{Rotate $\mathcal{P}_R$ around the vertex $D$ by $+\pi$ keeping fixed the identifications of the edges. This operation is equivalent to the rotation of $\mathcal{P}_R$ around its own center, followed by a translation by $-\mathbf{e}_1+\mathbf{e}_2$. Call this new domain $\mathcal{P'}$. The smectic field is invariant under translations along lattice vectors as well as rotations by $\pi$. This proves that the field is left invariant under this operation and so we conclude that the domains $\mathcal{P}$ and $\mathcal{P'}$ represent the same state.}
\par By discrete translational invariance of the layers along $\mathbf{e}_i$, we also conclude that we can translate an individual triangular face by an integer multiple of either basis vector without changing the state. This allows for a great simplification in the counting of states. We can consider an infinite plane tiled with domains $\mathcal{P}$. Using the symmetry operations described above, one concludes that the angles $\gamma\in(0,\pi/6) \ $,$(\pi/3,\pi/2)\ $, $(2\pi/3,5\pi/6)$ represent the same Left-handed states and therefore should not be over-counted. Analogously, the angles $\gamma\in(\pi/6,\pi/3)$,$(\pi/2,2\pi/3),(5\pi/6,\pi)$ describe the same states (Fig.~\ref{angles}).  {In other words, the angle $\gamma$ is defined $\mod \pi/3$, which is again commensurate with the angles $\pi/3$ contained in the equilateral faces.} In conclusion, we just need to consider a wedge of aperture $\pi/6$ in a triangular lattice, count the allowed states in the wedge, and multiply by 2 to account for the right-handed partners.
\par
\begin{figure}[b]
\includegraphics[height=5cm]{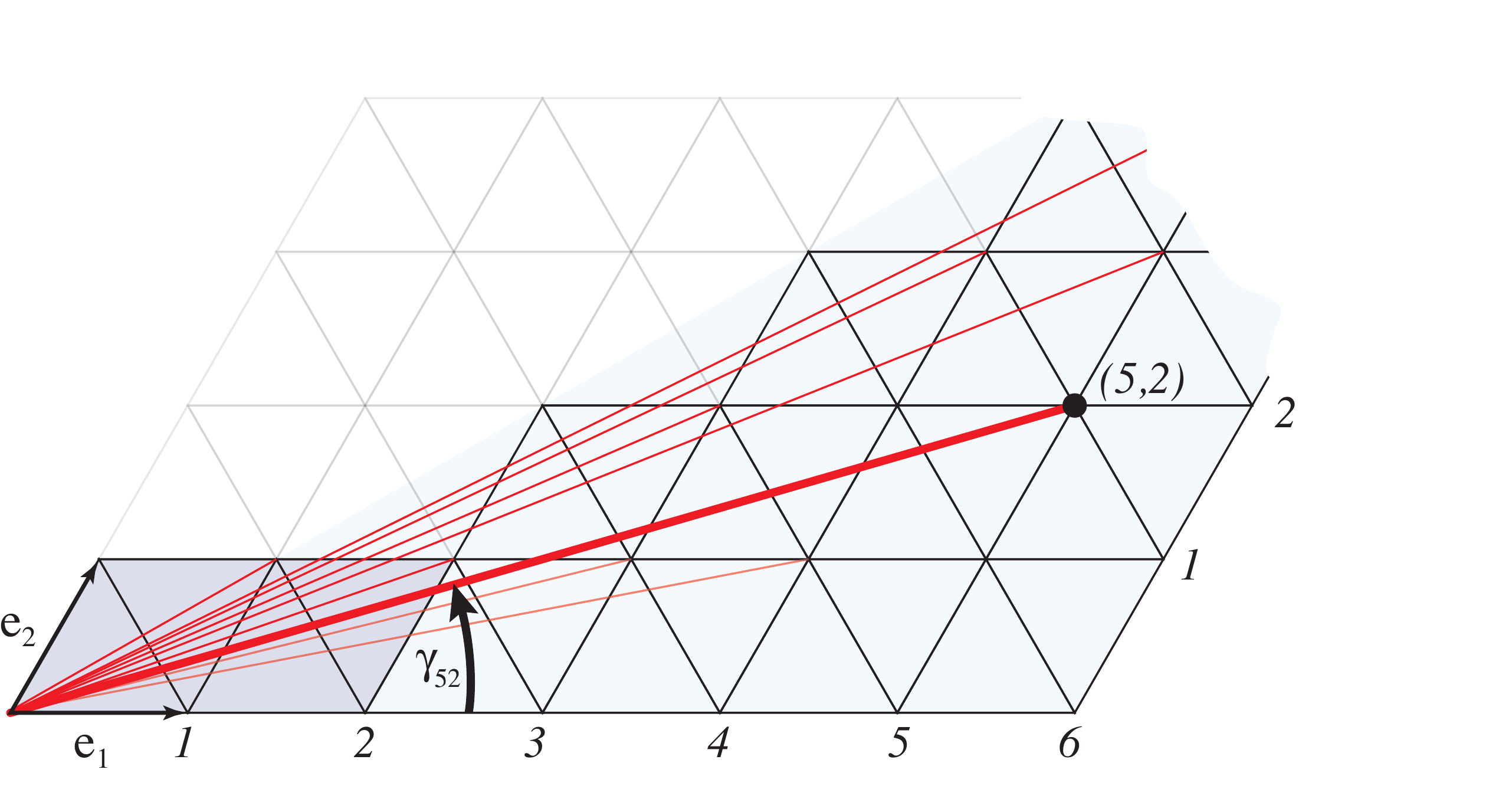}
\caption{\label{orchard}The Euclid's Orchard procedure to label the allowed smectic tilt angles $\gamma_{nm}$.}
\end{figure}

 At first, consider the states that contain two parallel spirals. The triangular lattice provides the optimal setting to classify the allowed angles. Rule (i) implies that we can find the allowed orientations by drawing rays that connect the origin of the lattice with appropriate lattice vertices in a $\pi/6$ wedge. We label the vertices by pairs of integers $(n,m)$ corresponding to multiples of the lattice directions $\mathbf{e}_1,\mathbf{e}_2$. The position of a vertex from the origin is $\mathbf{r}=n\mathbf{e}_1+m\mathbf{e}_2$, with $\mathbf{e}_1\cdot\mathbf{e}_2=1/2$. The allowed directions will be labeled by $\gamma_{nm}$. See Fig.~\ref{orchard}.
\par It is easy to see that, for a fixed ray starting from the origin, all those vertices that are intersected by the ray correspond to the same state. If we imagine an observer at the origin, then a state is labeled by the first vertex in his/her line of sight. In terms of the vertex labels, this means that a state is labeled by a pair $(n,m)$ where $m$ and $n$ are relative primes (or, the fraction $m/n$ is irreducible). This identifies the space of allowed angles as the so-called Euclid's Orchard \cite{0305-4470-25-1-005}. The labeling, and therefore the number of states, does not depend on the specific shape of the triangles, a result that is compatible with the degeneracy in the tetrahedral vesicle's shapes. 
\par The values of the angles, on the other hand, depend on the relative size of the edges. For a regular tetrahedron, we find:
\begin{align}
\label{eq:tangamma}
\tan\gamma_{nm}&=\frac{\sqrt[]{3}\, m}{2n+m} \quad, \text{ $(m,n)$ coprimes.}\\
\label{eq:sin}
\sin\gamma_{nm}&=\frac{\sqrt{3}\, m}{2\sqrt{n^2+m^2+nm} }\quad, \text{ $(m,n)$ coprimes.}
\end{align}
The result does not depend on the overall size of the tetrahedron's edge $L$, which we expect from the scale invariance of the problem. In addition, equations \eqref{eq:tangamma} and \eqref{eq:sin} depend only on the ratio $q\equiv m/n$, which is consistent with the state counting: non-irreducible fractions lead to the same angles.
\par Finally, according to rule (ii), we can insert an arbitrary number of equally spaced layers $\nu$ between the first and the second spiral (say between the vertices $A$ and $B$). Therefore, we conclude that a state $S_{m,n,\nu}(\alpha,r)$ is labeled by the integers $m,n,\nu\in \mathbb{N}$ where $m,n$ are relative primes and label the layers tilt angle, $\nu$ labels the number of \emph{independent} layers in the foliation \footnote{The number $\nu$ of independent layers is less than or equal to the number $N$ of parallel lines drawn on the developed tetrahedral shell.}, and $\alpha$ and $r$ are the angle between $\mathbf{e}_1$ and $\mathbf{e}_2$, and the ratio between their magnitudes. The latitudinal and quasi-baseball states occur for the special values $m=0$ and $m=n=1$ respectively.
\par {We will now describe the procedure to draw a smectic state on $\mathcal{P}$ that contains only two parallel spirals: (a) choose a vertex labeled by a pair of coprimes $(n,m)$ on the lattice and draw a line ${l}$ that joins the origin with $(n,m)$. (b) Occupy every other vertices in the lattice with lines parallel to $l$. (c) Choose any fundamental domain in the lattice and fold it.  According to this procedure, a  choice of $\gamma$ automatically fixes a layer spacing $h_\mathrm{max}$. Notice that the parameter $m$ is also the number of times that the edge $\mathbf{e}_1$ is intersected by the smectic layers, an observation that will be useful later. 
From the double-spiral state described above, we can form a new class of states by stacking additional layers between the spirals. We can insert an arbitrary number of them, provided that we keep the layers equally spaced (the layer spacing is reduced by an integer factor $p$: $h_\mathrm{max}\to h_\mathrm{max}/p$). 
\par In this way, we are increasing the density of smectic layers $\eta\equiv L/h\to p \eta$ on the surface of the tetrahedron by integer multiples.  The added layers are parallel to the spirals, but form closed loops, and thus constitute a new class of topological smectic states that is independent from the spirals and the latitudinal states. We will call this new class of smectic patterns as double spiral-loop (2SL) states. We will prove later that such a double-spiral loop state is described by a reducible fraction, where both $n$ and $m$ have been multiplied by the same integer $p$.}
\par If we shift any of the patterns described above by half layer spacing along the direction orthogonal to the layers, we obtain another class of states. The structure of the defects at each vertex is a chiral version the quasi baseball defect shifted by a half layer spacing. The single spiral state can be analyzed in a similar way.
\begin{figure}[t]
\includegraphics[width=8cm]{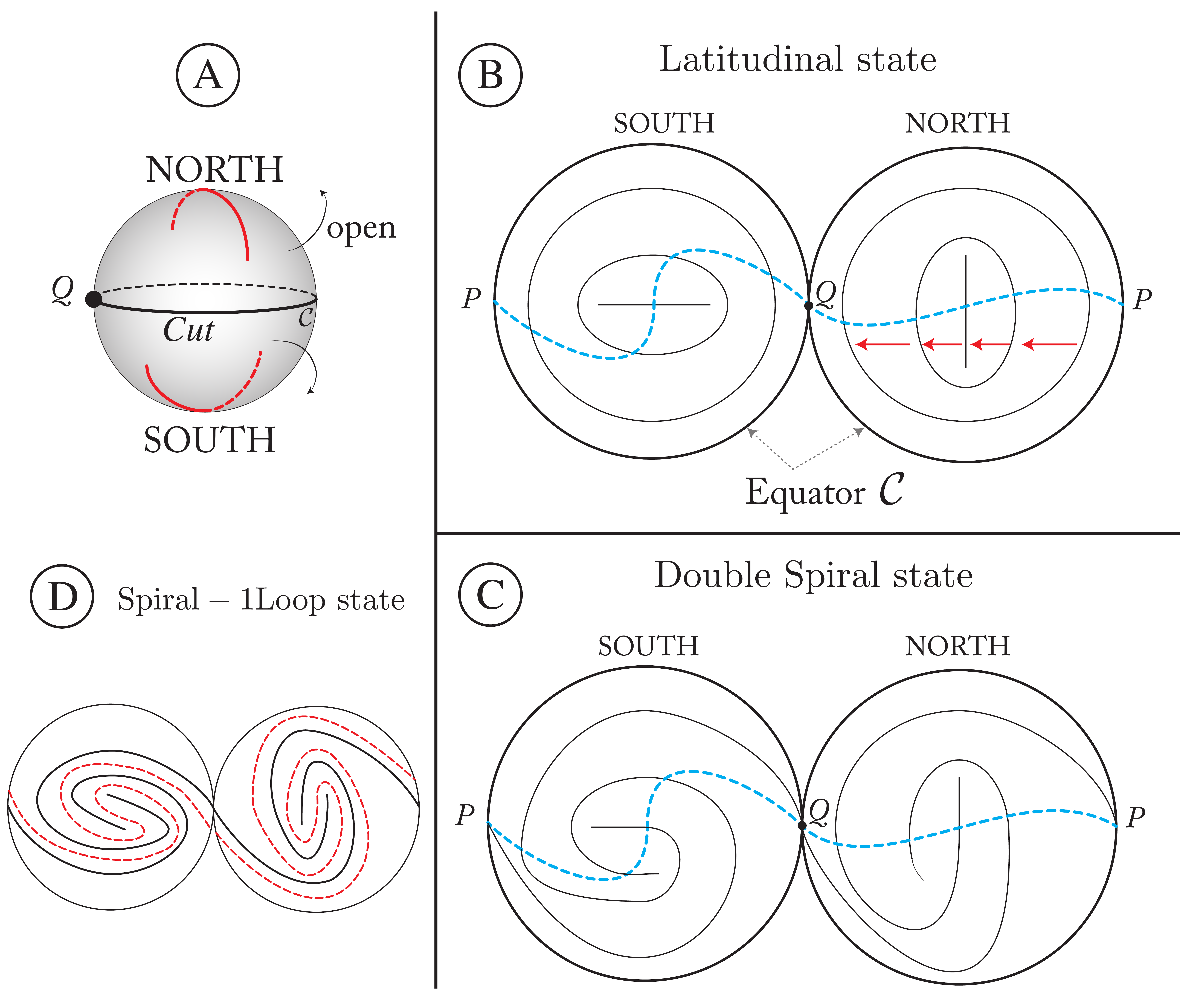}
\caption{\label{proofFIG} The figure proves that the latitudinal state (\emph{B})  and the double spiral state (\emph{C}) are topologically distinct. All the layers must be cut along the dotted line and shifted globally by one layer spacing. A shift by 2 layer spacings (\emph{D}) gives a state containing two spirals and a loop (red dotted line). }
\end{figure}
\section{Topology of the spiral states}
We will now prove that the pattern of the smectic layers on a spiral state is topologically distinct from the latitudinal state. 
Notice that distinguishing a single spiral from a latitudinal state requires more than the usual winding number $q$ around the defect location \cite{Xing2009,Mermin_RevModPhys.51.591}. In fact, if we enclose a defect core with a loop and we measure the rotation of the tangent vector $\mathbf{t}$ around the loop, we find that for both patterns $q=+1$. We must use another characterization to distinguish a spiral from a latitudinal pattern.
\par For the purpose of topological classification, we can imagine the smectic pattern as painted on a sphere. Start with a latitudinal state (Fig.~\ref{proofFIG}A) on the sphere, where the defects are located away from the equator (as it is in the tetrahedron case). Choose the great circle $\mathcal{C}
$ that is equidistant from all four defects and cut all along it except for a single point $Q$. Then, open the two hemispheres using the common point $Q$ as a hinge, and stereographically project the smectic pattern on the disks (one for each hemisphere) that have $\mathcal{C}$ as its boundary. Draw a line through the hemispheres that cuts every circular layer twice and cuts only once the layers connecting the vertices (see the blue dotted line in Fig.~\ref{proofFIG}B). Then cut the layers where they intersect the dotted curve and shift them by one layer spacing in the direction of the red arrows. After the layers are rejoined and the hemispheres have been glued again along $\mathcal{C}$, we find the double spiral state. 
\par This proves that the two states are topologically distinct, although they have the same topological charge $(4\times1/2)$. Penrose considered a similar discontinuous transformation between similar states in an infinite plane \cite{doi:10.1111/j.1469-1809.1979.tb00677.x}. This is a new manifestation of the fact that in smectic liquid crystals the disclination charge is not sufficient to give a 1:1 characterization of the states. 
If we shift the pattern by integer multiples of the layer spacing and reconnect them, we obtain the Spiral-Loop (SL) states. 
\par A single-spiral state on the sphere is parametrized by the curve $\sigma$:
\begin{equation}
\label{eq:1S}
\varphi(t)=p\,\theta(t) \quad, p\in\mathbb{N}
\end{equation}
where $\varphi,\theta$ are spherical coordinates on $S^2$, $\theta(t)=\pi t$ and $t\in[0,1]$. The integer $p$ is the number of times we cross the curve $\sigma$ if we travel from the north pole to the south pole along a great circle. The winding number of \eqref{eq:1S} around the North pole is +1.
The double spiral states can be described as bound pairs of $+1/2$ disclinations (see Fig.~\ref{classes}-III - the winding number around both vertex $A$ and $B$ is +$1/2$). 
\section{Degeneracy of states at fixed density}
In a physical realization of a smectic vesicle, there are two relevant length scales: the vesicle size $L$ (we will consider again regular tetrahedra) and the size of the molecule, which determines the preferred layer spacing $h$. These two length scales fix the density of layers on the vesicle through the ratio $\eta\equiv L/h$. 
{The parameter $\eta$ has a lower limit of $2$ (when $h=L/2$) and an upper limit of $L/a$, where $a$ is a microscopic cutoff, comparable with the molecular length scale. }
We wish to study the number of allowed states $S(\gamma_{nm})$ at fixed $\eta$. The parameter $\eta$ is related to the density of smectic layers in a vesicle of fixed size $L$.
\par Consider the plane with cartesian coordinates $x,y$ and a set $\mathscr{L}$ of parallel lines $y=kh$, where $k$ is an integer and $h$ is a fixed layer spacing. Draw a fundamental domain $\mathcal{P}$ spanned by 
\begin{equation}
\mathbf{e}_1=(L,0) \quad ,\quad\mathbf{e}_2=\frac{L}{2}(1,\sqrt[]{3})  \quad.
\end{equation}
In order to find the allowed angles at fixed $h/L$, we will rotate the domain $\mathcal{P}$ by $\gamma$ from $0$ to $-\pi/6$. By rule (i), an allowed state must have both vertices $B,D$ (See Fig.~\ref{fund_dom}) occupied by a layer. These two vertices have positions $\mathbf{e}_1$ and $\mathbf{w}\equiv\mathbf{e}_1+\mathbf{e}_2$. Rule (i) implies that the tip of the rotated vectors $R_\gamma\mathbf{e}_1$ and $R_\gamma\mathbf{w}$ must simultaneously intersect a line in the set $y=zh$. This condition is satisfied only if the angle $\gamma$ is the solution to~\eqref{eq:tangamma}. Finally, we know that for a given $\gamma_{nm}$, the layers $\mathscr{L}$ divide the edge $\mathbf{e}_1$ into $m$ segments of length $h/\sin\gamma$:
\begin{equation}
\label{eq:intersection}
L=\frac{mh}{\sin\gamma_{nm}} \quad, 
\end{equation}
{ Using \eqref{eq:sin} into \eqref{eq:intersection}, we can rewrite~\eqref{eq:intersection} in terms of $n,m$ to get the relation between the allowed angles and the density:}
\begin{equation}
\label{eq:degeneracy}
n^2+m^2+nm= \frac{3\eta^2}{4} \quad, \quad m<n  \quad,
\end{equation}
which is a form of Diophantine equation.
{The symmetry $n\leftrightarrow m$ of  \eqref{eq:degeneracy} exchanges the chirality of the spirals from left to right-handed. When $h\ll L$, we can stack a large number of layers on the tetrahedron's surface, and we expect to find many states with the same density $\eta$, labeled by different values of the angles. 
For example, the density $\eta=2 \,\sqrt[]{247}/(3\,\sqrt[]{3})$ allows two tilt angles $\gamma_{nm}$, one with $(n,m)=(11,7)$ or $\gamma\approx 22.7^\circ$, the other with $(\bar{n},\bar{m})=(14,3)$ or $\gamma\approx 9.5^\circ$.}
\par {Equation \eqref{eq:degeneracy} correctly describes SL states as well. In fact, multiplying equation \eqref{eq:degeneracy} by $p^2$ we find:
\begin{equation}
(pn)^2+(pm)^2+(pn)(pm)= \frac{3(p\eta)^2}{4} \quad, \quad p\in\mathbb{N} \quad,
\end{equation}
where $\bar{\eta}=p\eta$ is the new density. Notice that the angle $\gamma$ given by \eqref{eq:sin} depends only on the ratio $q=n/m$. The transformation $(n,m)\to(pn,pm)$ leaves $q$ invariant, so we proved that we can increase the density by integer multiples without changing the tilt angle $\gamma$, and that the Spiral-Loop states are obtained by rescaling the coprime pairs by an integer number.}
\par  Can a Double-Spiral and a Spiral-Loop state have the same layer spacing? Let $\eta>\bar{\eta}$ be the densities of two spiral states whose angles are $\gamma_{nm},\gamma_{\bar{n}\bar{m}}$ respectively. We increase the density $\bar{\eta}$ by an integer $p\neq1$, thus obtaining a SL state, and ask: can $p\bar{\eta}$ be equal to $\eta$? The answer is yes if and only if
\begin{equation}
\label{eq:mixing}
\frac{n^2+m^2+nm}{\bar{n}^2+\bar{m}^2+\bar{n}\bar{m}}=p^2 \quad (p\neq1) \quad,
\end{equation}
where $n,m$ and $\bar{n},\bar{m}$ are coprimes. The Spiral and Spiral-Loop states can have the same layer spacing if there exist solutions of \eqref{eq:mixing} in $\mathcal{O}\times\mathcal{O}$. The number of independent smectic layers that close into loops is $p-1$. For example, the Spiral state with tilt angle $(n,m)=(20,17)$ and the SL state with tilt angle $(\bar{n},\bar{m})=(4,1)$ and 6 loops solve \eqref{eq:mixing} with $p=7$, and in fact they have the same density $\eta=7\bar{\eta}=14\,\sqrt[]{7}$.
\par This is an important result, because it shows that each state is labeled by a rational number (the angles are still counted by the Euclid's Orchard) and that a fixed value of $\eta$ can contain states from different topological classes. 
\par{Commensurability implies that only special values of the density allow for the existence of ground states. From equation \eqref{eq:degeneracy} we see that the values of $\eta$ are constrained by the fact that $(n,m)$ are integers. For example, there is no state associated to $\eta=\sqrt[]{37}$ because for this value \eqref{eq:degeneracy} does not have solutions in the form of a pair of integers $(n,m)$. The number of states at fixed $\eta$ should be computed by
\begin{equation}
\label{eq:deg}
\Omega(\eta)=\sum_{\substack{(n,m)\in\mathbb{N}^2\\
m\leq n}}\delta\biggl(\eta-2\,\sqrt[]{\frac{n^2+m^2+nm}{3}}\biggl) \quad,
\end{equation}
where $\delta$ is the Dirac delta distribution and the sum is taken over pairs of natural numbers, to take into account the SL states. Equation \eqref{eq:deg} counts the number of solutions to \eqref{eq:degeneracy} as we sample the pairs $(m,n)$ in the infinite triangular wedge $m\leq n$.
\begin{figure}
\includegraphics[width=8cm]{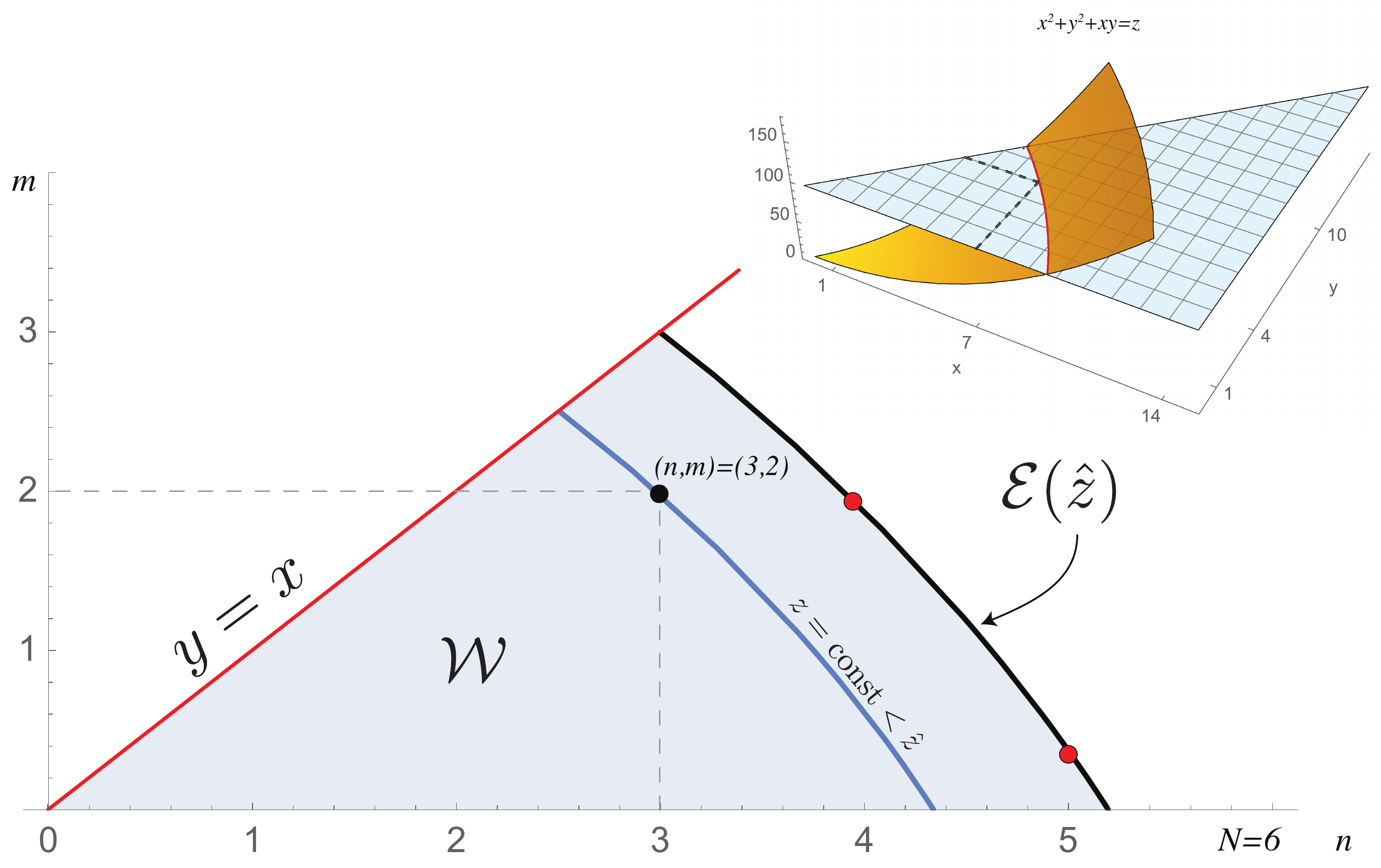}
\caption{\label{region} Sampling region for the numerical evaluation of $\Omega(z)$.}
\end{figure}
\par {We can solve numerically equation \eqref{eq:deg}. 
Instead of performing the sum \eqref{eq:deg}, it is convenient to change variables and look at the degeneracy as function of the auxiliary variable $z\equiv (3/4)\eta^2$. The result is unaffected by the change of variables, because all the quantities are positive definite, and the map $\eta\to\eta^2$ is 1:1. For each fixed $z$, we look at the intersection between the surface $h(x,y)=x^2+y^2+xy$ and the plane $h=z$ restricted to the region $y<x$ in $\mathbb{R}^2$. The two surfaces intersect along an arc of ellipse $\mathcal{E}$, as shown in Fig. \ref{region}. The number of states with density $\eta=\sqrt[]{(4z)/3}$ is given by the number of points $f_{nm}(z)$ on $\mathcal{E}(z)$ with integer coordinates. In order to sample the pairs $(n,m)$ correctly, we must sum over wedges $\mathcal{W}$ bounded by the $x$ axis, the line $x=y$, and by an ellipse $\mathcal{E}$. First, we introduce a cutoff $\hat{z}$ that parametrizes the elliptic edge of $\mathcal{W}$. This fixes an upper limit $\hat{n}$ on the sum over $n$. How large should $\hat{n}$ be in order to compute $\Omega(z)$ for all $z<\hat{z}$ ? At least, $\hat{n}$ should solve $\hat{z}=\hat{n}^2+\hat{n}+1$. For convenience, we choose $N$ to be the solution to $\hat{z}+1=\hat{n}^2+\hat{n}+1$:
\begin{equation}
\hat{n}=\mathrm{Int}\,\frac{1}{2}\big(\sqrt[]{1+4\hat{z}^2}-1\big) \quad,
\end{equation}
where $\mathrm{Int}$ is the integer part, and we keep in mind that the resulting $\Omega(z)$ is accurate only for $z<\hat{z}$.
The sum over $n$ and $m$ are not independent. As $n$ ranges from $1$ to $\hat{n}$, $m$ varies as
\begin{equation}
1<m<M(n)\equiv\frac{1}{2}\biggl(\sqrt[]{\frac{4\hat{z}^2-3n^2}{3}}-n\biggl) \quad.
\end{equation}
\par Finally, consider the single spiral states. If the tetrahedron contains a single spiral and no closed loops, we have
\begin{equation}
\label{eq:d1}
h=2L\sin\gamma_n
\end{equation}
where 
\begin{equation}
\label{eq:d2}
\sin\gamma_n=\frac{\sqrt[]{3}}{2 \ \sqrt[]{n^2+n+1}}
\end{equation}
where $n$ is the number of times that the spiral intersects the edge $L$ of the tetrahedron. Equation \eqref{eq:d2} can be obtained from \eqref{eq:sin} setting $m=1$. Using \ref{eq:d2} in \ref{eq:d1} we find
\begin{equation}
3\eta^2=n^2+n+1 \quad, \quad n\in\mathbb{N}
\end{equation}
Suppose we take a single spiral configuration on the developed tetrahedron.
We can insert an arbitrary \emph{even} number of equispaced lines below the spiral (see Fig.).
\begin{figure}
\includegraphics[width=8cm]{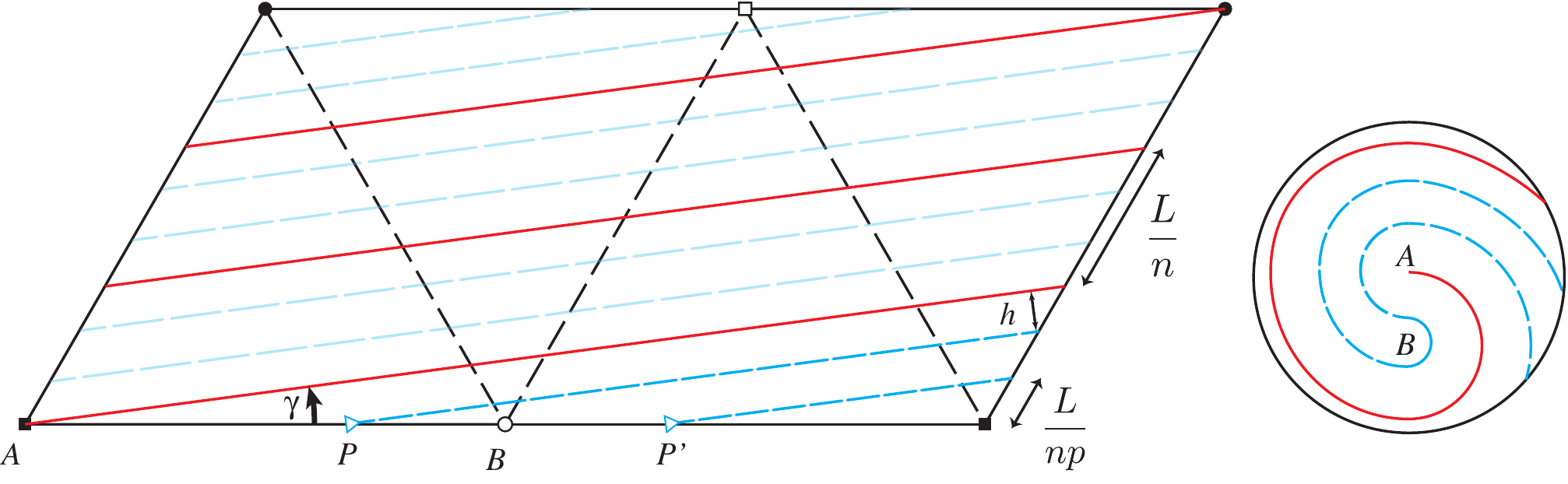}
\caption{\label{sinlgeSpiral}\emph{Left:} Construction of a  state containing a single spiral (red solid line) and one closed loop (blue dashed line). The points $P$ and $P'$ coincide when the domain is folded. \emph{Right:} The defects structure projected on a sphere, as seen from the South pole.}
\end{figure}
These will reconnect trivially across the edges. We have constructed a state with a single spiral and a certain number of loops. So we are free to increase the density by odd multiples. This leads to 
\begin{equation}
\label{eq:NUM4}
3\eta^2=p^2(n^2+n+1) \quad, \quad n\in\mathbb{N} \quad p \ \ \mathrm{odd}
\end{equation}
$p$ is odd because it measures the number of times that the segment $L/n$ is divided by the new layers.
These states are chiral so they have a weighting factor of 2 . If we shift the pattern by $h/2$ in the direction normal to the layers we get the same state again. As we discussed for the double spiral state, we can sample the pairs $(n,p)$ in a suitable region $\mathcal{W}'$ and collect the frequencies of $\eta$.
\par\section{Numerical evaluation of $\Omega(\eta)$}
For a fixed $\eta$, we want to count the number of allowed states. It is convenient to rewrite equations \eqref{eq:NUM1},\eqref{eq:degeneracy},\eqref{eq:NUM4} and $\eta=n$ (for the quasi-baseball state) in the following form:
\begin{align}
z&=4n^2 \quad , \quad n\in\mathbb{N}\\
z&=3n^2 \quad n\in\mathbb{N}\\
\label{eq:sampling1}
z&=p^2(n^2+n+1) \quad, \quad n\in\mathbb{N} \quad p \ \ \mathrm{odd} \\
\label{eq:sampling2}
z&=4(n^2+nm+m^2)\quad, \quad m<n\in\mathbb{N} 
\end{align}
where $z\equiv 3\eta^2$ and $n,m,p$ range in suitable domains. We have to count every density of the type \eqref{eq:sampling1} two times for chirality, and every density of type \eqref{eq:sampling2} four times - two for chirality and two for a rigid shift of the lines by $h/2$.
As we sample the integers $n,m,p$ in the appropriate domains, we collect the frequencies of $z$, and produce the set of pairs:
\begin{equation}
\biggl(\sqrt{\frac{z}{3}},\mathrm{freq}(z)\biggl)
\end{equation}
where $\sqrt[]{z/3}=\eta$.
\begin{figure}
\includegraphics[width=8cm]{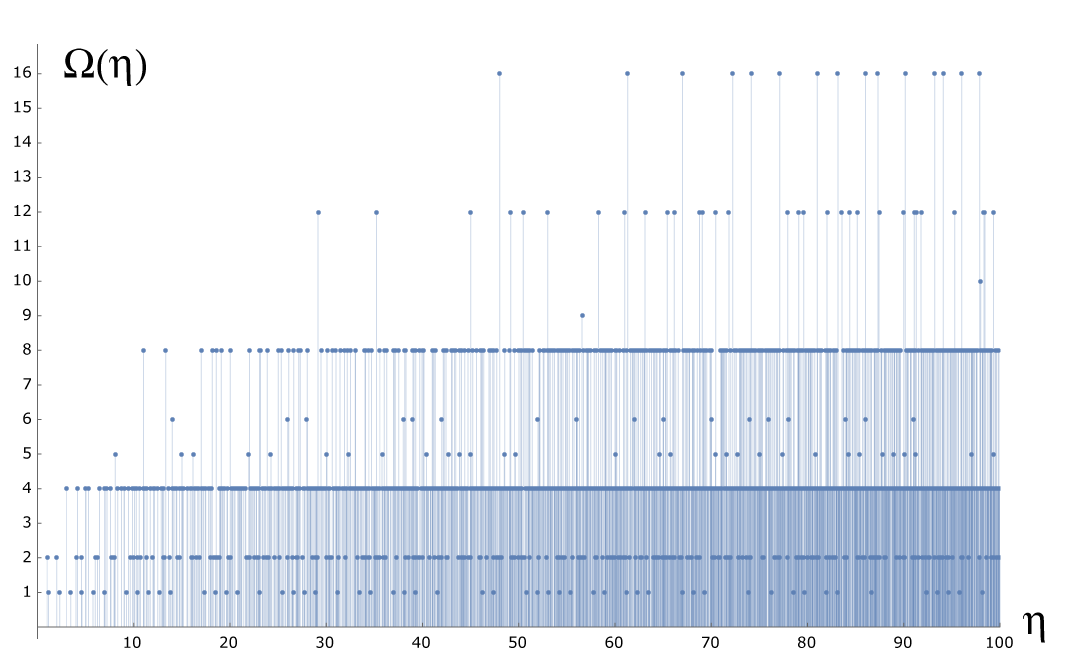}
\caption{\label{states} Degeneracy $\Omega(\eta)$ of tilted smectic states as function of $\eta\equiv\frac{L}{h}$ computed numerically for $\eta<100$.}
\end{figure}
The result for $\eta<100$,
is shown in Fig. \ref{states}. The degeneracy is a highly irregular function of the density. As $\eta$ increases, we meet higher and higher peaks, a result that we expect from \eqref{eq:degeneracy}, because the arclength of the ellipse grows and can allow more points $(n,m)$ on it (see also Fig. \ref{region}).
}
\section{Conclusions}
We studied the ground state ($T=0$) degeneracy of a vesicle made of smectic A liquid-crystalline building blocks. The shape of the vesicle and the configuration of the smectic layers  were inferred from an equilibrium model where the smectic order and the embedding of the shape are coupled in a Landau-De Gennes free energy. In order to find a solution to this infinite-dimensional minimization problem, we used the fact that the vesicle topology requires the presence of topological defects in the smectic layers. 
\par Focusing on the case of four $s=+1/2$ defects, we argued that at vanishing bending rigidity the vesicle's ground states are faceted tetrahedral shells and the layers must be parallel straight lines. The layers form topologically distinct patterns depending on the tilt angle $\gamma$ between the layers and an edge of the shell. We found that the system realizes all the topological defect configurations allowed by smectic order. Two of them, the latitudinal (i.e. $\gamma=0$) and the quasi-baseball (i.e. $\gamma=\pi/6$) states are topologically indistinguishable, but geometrically very different. In all other cases ($\gamma\neq 0,\pi/6$) the layers form either two parallel spirals or a combination of spirals and closed loops that wind around the shell. Because a spiral can be either right- or left- handed, these states reveal the breaking of chiral symmetry, an interesting result considering that the smectic A building blocks are achiral in nature.
\par It is possible that the vesicles could be functionalized by making use of the isotropic defect cores as preferred sites for ligand attachment. A single tetrahedral shell would then be approximated as a tetravalent unit \cite{2002NanoL...2.1125N}. The spiral states and their chiral nature might allow interesting possibilities for supramolecular chemistry.
\par The topology of the smectic layers should also affect the global material parameters of the vesicles. For example, if all layers form closed loops (zero tilt angle) the length of each loop cannot be deformed and molecules cannot easily flow between the layers.  We therefore expect the system to have a relatively large rigidity to an external compression.  When the state contains only spirals, a localized compression can propagate a long distance along the spiral. Thus, a vesicle composed of spiral state should be softer than one composed of latitudinal state because it is able to accommodate shape deformations more easily. Consequences of these properties should be experimentally observable. For example, we expect that the diffusion coefficient through a perforated membrane should increase as the concentration of spiral states over latitudinal states increases because vesicles with spiral states can squeeze trough the membrane pores more easily.

\section{Acknowledgments}
FS and MJB thank the Syracuse Soft and Living Matter Program for support and the KITP for the generous hospitality during the completion of some of this work. Work by MJB was supported by the KITP grant PHY-1748958.
SRN was supported by NSF MRSEC DMR-1420709 and by NSF DMR-1404841 and acknowledges the generous hospitality of KITP. 
\section{Author contribution statement}
All authors discussed the structure of topological defects in smectic vesicles. The potential influence of tilt on the structure of vesicles with internal smectic order originated in discussions between MJB and SRN.
FS performed theoretical calculations and numerical computations. MJB supervised the project. FS wrote the draft of the article. All authors discussed the results and contributed to the final manuscript.
%


\end{document}